\documentclass{ws-procs9x6}
\newcommand{\Lam}{\Lambda(1405) }
\usepackage{wrapft}
\input{epsf}

\begin{document}

\title{Chiral dynamics of the two $\Lambda(1405)$ states
}

\author{ D. Jido$^1$,  J.A. Oller$^2$, E. Oset$^3$, A. Ramos$^4$ and 
U.-G. Mei\ss{}ner$^5$
}

\address{$^1$ECT$^{*}$, Villa Tambosi, Strada delle Tabarelle 286, I-38050 
Villazzano, Italy\\
$^2$Departamento de F\'{\i}sica, Universidad de Murcia, 30071 
Murcia, Spain\\
$^3$Departamento de F\'{\i}sica Te\'orica and IFIC,
Centro Mixto Universidad de Valencia-CSIC, 
Aptd. 22085, 46071 Valencia, Spain\\
$^4$Departament d'Estructura i Constituents de la Mat\`eria,
Universitat de Barcelona, Diagonal 647, 08028 Barcelona, Spain\\
$^5$ HISKP, University of Bonn, Nu{\ss}alle 14-16, D-53115 Bonn, 
Germany 
}

\maketitle

\abstracts{
 The $\Lam$ resonance is studied by a chiral unitary approach, in
 which the resonance is dynamically generated in coupled-channel meson-baryon
 scattering. Investigating the analytic structure of the scattering amplitudes
 obtained by the chiral unitary approach, we find two poles around the energies
 of the $\Lam$ coupling differently to the meson-baryon
states.  We reach the conclusion that the $\Lam$ resonance seen in experiments is
not just one single resonance, but a superposition of these two states. }

The $\Lambda(1405)$ resonance has been a long-standing example of a
dynamically generated resonance appearing naturally in scattering theory with
coupled meson-baryon channels\cite{dalitz}. Modern chiral
formulations of the meson-baryon interaction within unitary frameworks all lead to
the generation of this resonance\cite{ChULam,oller,bennhold}. 
Yet, it was shown that in
some models one could obtain two poles close to the nominal $\Lambda(1405)$
resonance, as it was the case within the cloudy bag model\cite{fink}. 
Also, in the investigation of the poles of the scattering matrix within the
context of chiral dynamics\cite{oller}, it was found that
 there were two poles close to the nominal $\Lambda(1405)$ resonance both
 contributing to the $\pi\Sigma$ invariant mass distribution.  This
 was also the case in other works~[\refcite{twolam}].
In this paper, we summarize this important theoretical finding of the two pole
structure of $\Lam$ based on Ref.~\refcite{Jido:2003cb}.

The $\Lam$ resonance here is described as a dynamically generated object in
coupled-channel meson-baryon scattering with $S=-1$ and $I=0$
within the chiral unitary approach\cite{angels}.
Respecting the flavor SU(3) symmetry, we consider the octet mesons 
($\pi$, $K$, $\eta$) and the octet baryons ($N$, $\Lambda$, $\Sigma$, $\Xi$)
in the scattering channels. The unitary condition is imposed  by summing up
a series of relevant diagrams non-perturbatively in a way guided by the 
well-established procedures in the 60's, such as the N/D method, which are
generally expressed in complicated integral equations.  
The good advantage of our approach is to obtain an analytic solution of the
scattering equation under a low energy approximation in which one takes
only the $s$-channel unitarity and limits the model space of the unitary
integral to one meson and one baryon states\cite{oller}. This is essential to 
study the resonance structure in detail, since the resonance is expressed as a
pole of the scattering amplitude in the second Riemann sheet.
The details of the model are given in \mbox{Refs.~\refcite{oller,bennhold}}.

\begin{table}[tbh]
\tbl{Pole positions and couplings to $I=0$
physical states from Ref.~4.
 \label{tab:jido0}}
{\footnotesize
\begin{tabular}{ccccccc}
\hline
 $z_{R}$ & \multicolumn{2}{c}{$1390 - 66i$} &
\multicolumn{2}{c}{$1426 - 16i$} &
 \multicolumn{2}{c}{$1680 - 20i$}  \\
 & $g_i$ & $|g_i|$ & $g_i$ & $|g_i|$ & $g_i$ & $|g_i|$ \\
 \hline
 $\pi \Sigma$ & $-2.5+1.5i$ & 2.9 & $0.42+1.4i$ & 1.5 & $-0.003+0.27i$ &
 0.27 \\
 ${\bar K} N$ & $1.2-1.7i$ & 2.1 & $-2.5-0.94i$ & 2.7 & $0.30-0.71i$ &
 0.77 \\
 $\eta\Lambda$ & $0.01-0.77i$ & 0.77 & $-1.4-0.21i$ & 1.4 & $-1.1+0.12i$ &
 1.1 \\
 $K\Xi$ & $-0.45+0.41i$ & 0.61 & $0.11+0.33i$ & 0.35 & $3.4-0.14i$ &
 3.5 \\
 \hline
 \end{tabular}}
\end{table}

Shown in Table~\ref{tab:jido0} are the positions of the poles in the second
Riemann sheet of the scattering amplitude with $S=-1$ and $I=0$ obtained by the
chiral unitary approach\cite{bennhold}. The coupling strengths of the resonances
to the meson-baryon states are also obtained as the residues of the
amplitude at the pole position.
We see that there are two poles around the energies of the
$\Lam$ showing a different nature of the coupling strength: the lower
resonance strongly couples to the $\pi\Sigma$ state, while the higher pole
dominantly couples to the $\bar KN$ state.

 \begin{figure}[bth]
    \epsfxsize=10cm
    \centerline{\epsfbox{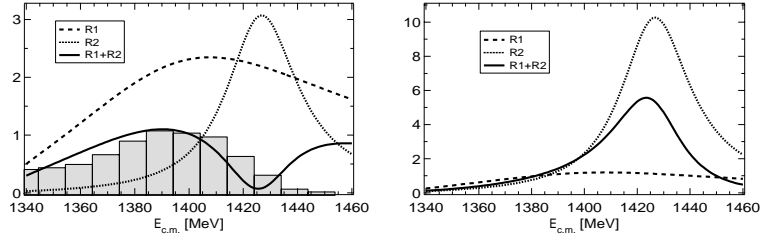}}
\caption[1]{The $\pi\Sigma$ mass distributions calculated from the toy model 
in Eq.~(\ref{eq:pisigamp}) for  $\pi\Sigma \rightarrow \pi\Sigma$
(left panel) and Eq.~(\ref{eq:kbarNamp}) for $\bar{K}N\rightarrow  \pi\Sigma$
(right panel). The dashed, dotted and solid lines denote the
contributions from the first term, the second term and the coherent sum of the
two terms, respectively.  The histogram in the left panel shows experimental
data\cite{hist}. Units are arbitrary. 
\label{fig:spec}}
\end{figure}

Let us see how these two poles appear in the physical observable using a toy
model in which amplitudes are described 
by the sum of two Breit-Wigner formulae corresponding to two resonances,
$R_{1}$ and $R_{2}$, such that:
\begin{eqnarray}
&& g_{\pi\Sigma}^{R_{1}} \frac{1}{W - M_{R_{1}} + i\Gamma_{R_{1}}/2}
        g_{\pi\Sigma}^{R_{1}} +
      g_{\pi\Sigma}^{R_{2}} \frac{1}{W - M_{R_{2}} + i\Gamma_{R_{2}}/2}
        g_{\pi\Sigma}^{R_{2}} \,\, ,
        \label{eq:pisigamp}\\
&& g_{\bar{K}N}^{R_{1}} \frac{1}{W - M_{R_{1}} + i\Gamma_{R_{1}}/2}
          g_{\pi\Sigma}^{R_{1}} +
      g_{\bar{K}N}^{R_{2}} \frac{1}{W - M_{R_{2}} + i\Gamma_{R_{2}}/2}
          g_{\pi\Sigma}^{R_{2}}\ ,
          \label{eq:kbarNamp}
\end{eqnarray}
where the resonance parameters have been taken from Table~\ref{tab:jido0}.
The former amplitude corresponds to the process $\pi\Sigma
\rightarrow \pi \Sigma$ and the later does to $\bar{K}N
\rightarrow \pi\Sigma$.
Shown in Fig.~\ref{fig:spec} is the modulus
square of these two amplitudes multiplied by the $\pi\Sigma$
momentum as a function of the energy. We also show the
contribution of each resonance by itself (dotted and dashed
lines). In both cases 
only one resonant shape (solid line) is seen, 
but the simulated $T_{\pi\Sigma \rightarrow \pi\Sigma}$
amplitude in the left panel of Fig.~\ref{fig:spec} produces a resonance at a
lower energy and with a larger width. This case reproduces very well the
nominal experimental $\Lambda(1405)$. 
However, if the invariant mass distribution of
the $\pi\Sigma$ states were dominated by the $\bar{K}N \rightarrow \pi\Sigma$
amplitude, then the second resonance $R_{2}$ would be weighted
more, since it has a stronger coupling to the $\bar{K}N$ state,
resulting into an apparent narrower resonance peaking at higher
energies as shown in the right panel~of~Fig.~\ref{fig:spec}.

\begin{wrapfigure}{r}{5cm}
\epsfxsize=5cm
\centerline{\epsfbox{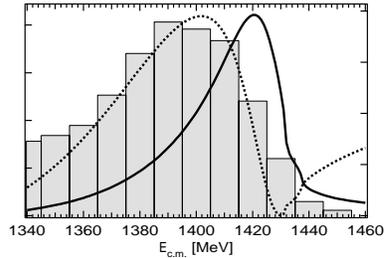}}
\caption[2]{The $\pi\Sigma$ mass distributions with $I=0$ constructed from
  the $\pi\Sigma \rightarrow \pi\Sigma$ (dotted lile) and $\bar{K}N\rightarrow
\pi\Sigma$ (solid line) amplitudes obtained by the chiral unitary approach. 
The histogram shows experimental
data\cite{hist}.  Units are arbitrary.
\label{fig:massdist}}
\end{wrapfigure}

The existence of the two pole is strongly related to the flavor 
symmetry. The underlying SU(3) structure of the chiral Lagrangians
implies that 
a singlet and two octets of dynamically generated resonances should appear, 
to which the $\Lambda(1670)$ and the $\Sigma(1620)$ would 
belong\cite{bennhold},
and that the two octets get degenerate in the case of exact SU(3) symmetry.  In
the physical limit, the SU(3) breaking resolves the degeneracy of the
octets, and, as a consequence, one of them appears quite
close to the singlet pole around energies of the $\Lam$ resonance.

The double pole structure of $\Lam$ found here should be confirmed by new
experiments. Clearly a reaction which forces the initial channel to be $\bar K N$
produces a different distribution with a narrower peak at higher energy than the
original distribution observed in the $\pi\Sigma \rightarrow \pi\Sigma$ channel,
since the former reaction gives more weight to the second resonance as shown in
Fig.~\ref{fig:massdist}, 
where we show the $\pi\Sigma$ mass distributions with
$I=0$ initiated by the $\pi\Sigma$ (dotted line) and $\bar KN$ (solid line)
states in the chiral unitary approach\cite{bennhold}.
One problem here is that one cannot access the second resonance directly from
the $\bar K N$ scattering, since the resonance lies below the threshold of the
$\bar K N$ state. Therefore one has to lose energy of the $\bar K N$ state before
the creation of the resonance. 
One possibility is to have the $\bar{K}$ lose some energy by emitting a 
photon, as done 
in Ref.~\refcite{Nacher:1999ni}
in the study of the $K^- p \to \gamma \Lambda(1405)$
reaction. 
Another possibility is provided in Ref.~\refcite{Hyodo:2004vt}, where
the photo-induced $K^{*}$ production on proton has been discussed,
and this process has been found suitable to isolate the second
resonance.

In conclusion, the chiral unitary approach suggests that two resonances are
dynamically generated around energies of the nominal $\Lam$. Since
they are located very close to each other, what one sees in experiments is a
superposition of these two states. 
The existence of the two poles can be found out by performing different
experiments of the creation of the $\Lam$ initiated by the $\bar
K N$ state. If one could confirm the double pole structure, it 
would be one of the strong indications that the structure of the $\Lam$ is
largely dominated by a quasibound meson-baryon component.

\end{document}